\documentclass[prl,twocolumn,showpacs]{revtex4}
\usepackage{array,amsmath,amssymb,graphicx}
\newcount\minute
\newcount\hour
\newcount\hourMins
\def\now{
\minute=\time
\hour=\time \divide \hour by 60
\hourMins=\hour \multiply\hourMins by 60
\advance\minute by -\hourMins
\zeroPadTwo{\the\hour}:\zeroPadTwo{\the\minute}
}
\def\zeroPadTwo#1%
{%
\ifnum #1<10 0\fi
#1%
}

\begin{document}
\newcommand{\bnabla}{\bm{\nabla}}
\newcommand{\bz}{{\mathbf z}}
\newcommand{\bx}{{\mathbf x}}
\newcommand{\br}{{\mathbf r}}
\newcommand{\bG}{{\mathbf G}}
\newcommand{\bu}{{\mathbf u}}
\newcommand{\bq}{{\mathbf q}}
\newcommand{\cH}{{\cal H}}

\title{\textsc{Order and Creep in Flux Lattices and CDWs Pinned by Planar Defects }}

\author{Aleksandra Petkovi\' c and Thomas Nattermann
}

\affiliation{Institut f\"ur Theoretische Physik, Universit\"at zu
K\"oln, Z\"ulpicher Str. 77, 50937 K\"oln, Germany}

\date{\today,\now}

\begin{abstract}
The influence of randomly distributed
point impurities \emph{and} planar defects
on the  order and transport in type-II
superconductors and related systems is
considered theoretically. For random planar
defects of identical orientation the flux
line lattice exhibits a new glassy phase
with
diverging shear and tilt modulus, a transverse
Meissner effect, large sample to sample
fluctuations of the susceptibility and an
exponential decay of translational long
range order. The  flux creep resistivity
for currents $J$ parallel to the defects
is $\rho(J)\sim \exp-(J_0/J)^{\mu}$  with $\mu=3/2$.
Strong disorder enforces an array   of
dislocations to relax shear strain.
\end{abstract}

\maketitle

\emph{Introduction}. Type-II
superconductors can be  penetrated by an
external magnetic field  in the form of
quantized magnetic flux lines (FL). Under
the influence of a transport current $ J$ FLs
will move and hence give rise  to
dissipation. The resulting linear
resistivity
is proportional to the magnetic induction
$\textbf{B}$  \cite{Bardeen+65}.
To stabilize superconductivity it is
therefore essential to pin FLs. One source
of pinning is point disorder. In
high-T$_c$ materials point disorder is
practically always existing because of the
non-stoichiometric composition of most
materials.  Then the system regains
superconductivity in the sense that the
linear resistivity vanishes
\cite{Blatter_93}. However, thermal
fluctuations lead to flux creep resulting
in a non-zero \emph{nonlinear} resistivity
of the form
$\rho(J)\sim
e^{-({J_P/J})^{\mu}}
$  where  $\mu=1/2 $ \cite{Na90}  . 
$J_P (\gg J)$ is a function of $B$, temperature
$T$ and the concentration and strength of
the pinning centers. This response of the
system on an external current is closely
related to the order of the FLL in the
presence of point pinning centers, which
shows a power law decay of its
translational order parameter in the
''Bragg glass'' phase
\cite{Na90,GiDo94,Klein}. Although
the linear conductivity is now zero, there
is still a finite resistivity for a finite
current. It is therefore indicated to
look for a more effective pinning
mechanisms corresponding to larger values of the 
creep exponent $\mu$.  One option is columnar defects
which lead to a  ''Bose glass'' phase with stronger pinning properties
(see e.g.
\cite{NeVi_93}). 
\begin{figure}[htbp]
\begin{center}
\includegraphics[width=0.8\linewidth]{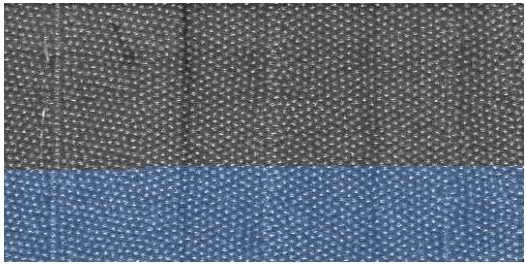}
\caption{\small{Planar crystalographic defects in BSCCO (bright vertical lines), unpublished. Figure by courtesy of M. Menghini, Y. Fasano and F. de la Cruz from Centro Atomico Bariloche, Argentina and Eli Zeldov from  Weizmann Institute, Israel.}}
\end{center}
\vspace{-0.2cm}
\end{figure}

An even more pronounced
effect can be expected from planar
defects like twin planes or grain
boundaries, which will be considered in
the present paper. Twins are ubiquitous in superconducting
YBCO and La$_2$CuO$_4$ where they are needed to accomodate strains arising from tetragonal to rhombic transformations. 
But also other causes are possible (see Fig.1). Planar defects occur
frequently in families with the same
orientation but random distances
\cite{many_parallel_twins,Paco}   or in
orthogonal families of lamella
(''colonies'')
\cite{many_orthogonal_twins}. The  mean
distance $\ell_D$ of the defect planes is
of the order of $10$ nm
\cite{many_parallel_twins} to $\mu m$
\cite{Dolan_89}. Pinning of individual FLs
by planar defects has been investigated in
the past both for clean and disordered
systems \cite{Blatter_93, BaKa94,marchetti}.
Recently it was shown that, depending
on the mutual orientation of the FL lattice (FLL) and
the defects,  dilute planar defects are
indeed  a relevant perturbation even in
the presence of point disorder
\cite{EmNa06}, provided they are parallel
to the main lattice planes of the FLL.
In systems with parallel
defect planes this is the generic
situation since the FLL will rotate in
such a position to reach maximum overlap
with the defects (provided $\textbf{B}$ is
aligned with the defect planes).
\begin{figure}[htbp]
\begin{center}
\includegraphics[width=1\linewidth]{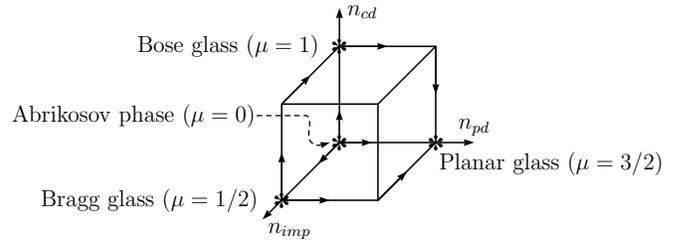}
\caption{\small{Disordered vortex lattices resulting from
impurities, columnar and planar defects of concentration $n_{imp}, n_{cd}$
and $n_{pd}$, respectively. In the presence  of planar defects
the planar glass phase is ultimately stable. $\mu$ denotes the creep exponent. }}
\end{center}
\end{figure}
\vspace{-0.5cm}
It turns out that the Bragg and the Bose glass phase are unstable with respect to the presence of 
many random  planar defects and will be substituted by a new type of a \emph{ planar glass phase} (see Fig. 2). In the present paper we will discuss the nature of this  phase.   The latter is characterized by a complex energy landscape  with many metastable states and diverging energy barriers leading to a new creep law with $\mu=3/2$,  large sample to sample fluctuations of the magnetic susceptibility, an   exponential supression of translational order in the direction perpendicular to the defects, a resistance against shear deformations as well as the occurrence of a transverse Meissner effect. 
If only displacements perpendicular to the defects are considered, as in the main part of this
 paper, our results
apply also to a wide class of systems which exhibit regular lattices of domain walls
 like magnetic slabs, charge density waves \cite{gruener} and
incommensurate systems \cite{BrCo78}.

\emph{Model}. We consider an Abrikosov FLL
in the presence of  randomly
distributed point impurities and random
defect planes, aligned with the magnetic
field. Since in both types of
imperfections superconductivity is
suppressed  they will attract FLs. Then
the Hamiltonian  reads \cite{Kogan_89}
\vspace{-0.15cm}
\begin{eqnarray}
 {\cal H}=\int
d^3r \frac{1}{2}\Big\{\sum_{\alpha\beta\gamma\delta}c_{\alpha\beta\gamma\delta}
(\partial_{\alpha} u_{\beta})(\partial_{\gamma}u_{\delta})+ \sum_{\alpha}{c_{44}^{(\alpha)}}(\partial_z u_{\alpha})^2\nonumber \\
+2\big[V_{\rm P}(\textbf{r})+ V_{\rm
D}(\textbf{r})\big]\rho(\textbf{u},\textbf{r})\Big\}  \,\,\,\,\, \,\,\,\,\,\,\,\, \,\,\,
\end{eqnarray}
where $\alpha,\beta,\gamma,\delta$ run over $x,y$. ${\bf u({\bf r})}=(u_x,u_y)$ denotes the FL displacement.
Only components of the elastic constants $c_{\alpha\beta\gamma\delta}$ with pairwise
equal indices are non-zero  \cite{dispersion}.
 $\rho(\textbf{u},\mathbf{r})=\rho_0\left\{-
{\bf \nabla}_{\perp}{\bf u}+
\sum_{\mathbf{G}}e^{i\mathbf{G}\left(\mathbf{r}_{\perp}-\mathbf{u}\right)}
\right\}\equiv \rho_s+\rho_p
$ is the FLL  density with $\rho_0=B_0/\phi_0$,
$\phi_0$ is the flux quantum.
$\textbf{G}$ is a reciprocal lattice
vector of the FLL and $\br_{\perp}=(x,y)$.
$V_P(\textbf{r})
$   denotes the
pinning potential resulting from randomly
distributed point impurities. We will
first consider the (realistic) case that
all defect planes have the same
orientation but random distances
\cite{many_parallel_twins}. Then the FLL
will orient itself such that its main
lattice planes will be parallel to the
planar defects to allow for  their maximal
overlap \cite{EmNa06}. The defect pinning
potentials  then has the form $
V_D(\textbf{r})=
-v_D\sum_{\textrm{d}}\delta({x}-{x}_{\textrm{d}})$
\cite{Blatter_93} where we assumed that
the defect planes are parallel to the
$yz-$plane. The $\delta$-functions are
considered to have a finite width of the
order of the  superconductor coherence
length $\xi_c$. A rough estimate for the
defect strength is given by $v_D\approx
H_{c}^2\xi_c^3$,
$H_c$ is the
thermodynamic critical field. 
The
statistical properties of the pinning
energies are then encoded in their pair
correlation functions $R_P({\bf u})$ and $R_D(u_x)$
for point disorder and planar defects, respectively.
Since the FLL
density includes a slowly varying and a
periodic part, $\rho_s$ and $\rho_p$,
respectively, we decompose the pinning
energy density accordingly. From the
periodic part we get 
 $R_D(u_x)=
(v_D\rho_0)^2/l_D\sum_{n\neq0}e^{i n 2\pi u_x/\ell}$, 
 $n$  is integer
\cite{footnote}.
$\ell \ll
\ell_D$, where $\ell$ and $\ell_D$ are the mean spacing between the
FLs and the defect planes, respectively.
The contributions from $\rho_s$ do not
contribute to the glassy properties of the
system since they can be eliminated by a
simple transformation \cite{Hwa94}.

Since our
main concern are the defect planes, it
seems to be  justified to start with a
simplified model in which only the
displacements $u_x\equiv u$ of the FLs
\emph {perpendicular} to the defect planes are
considered. Then only the elastic terms  with the coefficients $c_{xxxx}\equiv c_{11}$, $c_{yxyx}\equiv c_{66}$ and $c_{44}^{(x)}\equiv c_{44}$ remain in the Hamiltonian.  
From a
technical point of view it is convenient
to consider a generalization of our model
in $d$ dimensions by replacing $x$ by a
$(d-2)$-dimensional vector $\textbf{x}$.

\emph{Weak disorder }.
In the absence of defect planes point
impurities are relevant in less than $4$
dimensions.
The FLL
exhibits a phase with quasi long range
order : the Bragg glass
\cite{Na90,GiDo94,Klein},  which
exhibits a power law decay of  $ S_{\bf
G}({\bf r})=\overline{\langle
e^{i\textbf{G}(\textbf{u}(\textbf{r})-\textbf{u}(\textbf{0}))}\rangle}\sim
|\textbf{r}|^{-(4-d)}$.  The Fourier
transform of $S_{\bf G}({\bf r})$ is the
structure factor which has Bragg peaks.

It was recently shown in \cite{EmNa06} that dilute
planar defects can be a relevant
perturbation also in the presence of point
disorder.
Indeed, distorting the initially ordered FLL in volume $L^{d-2}L_yL_z$ the energy gain is of the order
$-(R_D''''(0)L^{d-2})^{1/2}L_zL_y$
whereas the elastic energy loss is
$c_{11}L_zL_yL^{d-4}$ since
distortions are  aligned
parallel to the defects. For $L\gg
L_D\sim
(c_{11}^2/R_D''''(0))^{1/(6-d)}$ the pinning energy gain wins and
the FLL starts to disorder in the
directions perpendicular to the defects.
The critical dimension below which weak
planar defects are relevant is $d=6$. 

For a more detailed study we use now a
functional renormalization group approach
 in $d=6-\epsilon$ dimensions. We follow
closely a related approach for columnar
disorder \cite{Balents,Fedorenko} but keep the
unrescaled quantities which correspond to
the effective parameters measured on scale
$L$. To lowest order the flow equations
for $\epsilon\ll 1$ read
\vspace{-0.15cm}
\begin{align}\label{eq:RG}
&{d\ln c_{ii}}/{d\ln L}=
2R_D''''(0)L^{\epsilon}/{(4\pi c_{11})^2},
\quad i=4,6\\\nonumber
&{dR_D(u)}/{d\ln L} =
R_D''(u)L^{\epsilon}\big(R_D''(u)-2R_D''(0)\big)/{(4\pi
c_{11})^2}.
\end{align}
Thermal fluctuations  and point
disorder are irrelevant for $\epsilon<4$ and $\epsilon<2$, respectively. There is no renormalization of
$c_{11}$ because of a statistical tilt
symmetry \cite{Schultz}. For $L\to L_D$, many metastable states appear and 
$R_{D}''(0)$ develops a slope discontinuity at
the origin which results in diverging elastic constants $c_{44}$ and $c_{66}$.
The renormalization can however be continued to $L\gg L_D$ if one imposes a small but \emph{finite
tilt} of the FLL such that $R''''(0)$ has to be replaced by $R''''(0^+)$ in Eq.~(\ref{eq:RG}).
In this case $c_{44}$ and $c_{66}$ remain finite but
 new terms of the form
$
\int_0^{2\pi}d\phi\left|{\Sigma_y}\cos\phi\,(\partial_{y}u)+
 {\Sigma_z}\sin\phi\,(\partial_{z}u)\right|/4\ell$
 are generated in the energy density which dominate the energy for small $u$.
The fixed point function
$R_{D}^{*''}(u,L)L^{\epsilon}=(2\pi c_{11})^2{\epsilon}
 \left[\frac{\ell^2}{36}-\frac{1}{3}\left(u-\frac{\ell}{2}\right)^2\right] $ for $0\le u<\ell$
is periodic in $u$  with
period $\ell$.
 The  newly generated terms renormalize according to
\vspace{-0.15cm}
 \begin{align}\label{eq:sigma2}
c_{66}^{-1/2}{d\Sigma_y}/{d\ln L}\approx c_{44}^{-1/2} {d\Sigma_z}/{d\ln L} \approx{\epsilon{\sqrt {c_{11}}}\ell^2}/{12
 L}.
\end{align}
$\Sigma_{z(y)}$ has the meaning of a interface
tension of a domain wall parallel to the
$\textbf{x}$ and $y$ ($z$) axes.
$\Sigma_z$ can be measured  by changing
the external magnetic field
by
$ H_x \mathbf{ \hat x}$
which  changes the Hamiltonian by
$-({B_0
}/{4\pi})\int d^3r\;
 H_x\partial_{z}u $.
To tilt the flux lines with respect to the
$z$-axis,  $ H_x$
has to overcome
the interface energy  $\sim\Sigma_z$  which
results in a threshold field  $ H_{x,c}={8\pi\Sigma_z\ell}/({\phi_0
\sqrt{3}})$ below which FLs remain locked
parallel to the planes. This is the
\emph{transverse Meissner effect}: a weak
transverse magnetic field $ H_x$ is screened from
the sample.  In this case  $c_{44}$ is infinite!
Only for $ H_x> H_{x,c}$ the average tilt of the FLs becomes  non-zero and $c_{44}$ stays finite.
Moreover,  there is a \emph{resistance against shear}   of the FLL:
the  shear deformation $\partial_yu_x$ is non zero (and $c_{66}$ finite)  only if
the shear stress $\sigma_{xy}$ is larger then a critical value $\Sigma_y/\ell$,
otherwise $c_{66}$ is infinite.  The divergence of  $c_{66}$
is a new property which does not exist in Bose glass.

An infinitesimal  change $\delta H_z\mathbf{\hat z}$ in the longitudinal
field allows to measure the longitudinal susceptibility
 {$\chi={B_0}\partial\langle\partial_x u \rangle/\partial H_z$.
The disorder averaged susceptibility
$\overline{\chi}={B_0^2}/4\pi c_{11}$ is independent
of the disorder as a result of the
statistical tilt symmetry.} The glassy
properties of the systems can most easily
be seen by the sample to sample
fluctuations of the magnetic
susceptibility
$\overline{\chi^2}-{\overline \chi}^2$.
Perturbation theory gives
$
(\overline{\chi^2}-{\overline
\chi}^2)/{\overline
\chi}^2=R_D''''(0)L^{\epsilon}/(5c_{11}^2)\sim
(L/L_D)^{\epsilon} ,
 $ %
 i.e. the sample to sample fluctuations of
 the susceptibility
grow with the the scale $L\lesssim L_D$,
$d<6$ which is a signature of a glassy
phase \cite{Hwa94}.

The structural correlations in this
phase are obtained in the standard way from $R_{D}''(u,L)$
\cite{Na90}  which gives $S_{\bf G}({\bf x},y,z) \sim |{\bf x}|^{-(6-d)}$.
In $d\le 4$ dimensions also the part of
the pinning potential related to $\rho_s$
becomes relevant which gives the
dominating contribution to the FL
displacements. Both, a Flory argument
\cite{Na90} and more detailed
calculations for a related one dimensional problem \cite{ViFE84,Fe80,GlNa04}
give in $d=3$
dimensions $S_{\bf G}({x},y,z) \sim e^{-|x|/L_D}$.
In the related study
\cite{ViFE84} Villain and Fernandez found from a non-perturbative RG
that for $d\le 4$ the disorder
renormalizes to strong coupling. We will show below that this case
gives qualitatively the same results.

To get more information about a real
$3$-dimensional system we consider next the
stability of this glassy phase  with
respect to point disorder by using an Imry-Ma  argument \cite{ImMa}.
The energy gain
from the point disorder in a region $L^{d-2}L_yL_z$  is  of the order $
-(\langle R_P(u)\rangle L_yL_zL^{d-2})^{1/2}$ \cite{footnote0} which has to be compared with the elastic energy loss
$L^{d-2}(c_{11}\ell^2L_yL_z/L^2+\Sigma_zL_y+\Sigma_yL_z)\sim L^d$.
If one ignores the fluctuations of $u$ and replaces $\langle R_P(u)\rangle $  by a constant
one finds that point disorder is irrelevant above $d=2$ dimensions. This critical dimension is further decreased
to zero if
the fluctuations of $u$ are taken into account by using $R_P(u)\sim L^{(d-6)/2}$.
A similar argument shows the irrelevance of columnar disorder.
  This argument
applies  for $L\gg L_D$ where $\Sigma_{y/z}$
has developed.

\emph{Flux creep.} Next we consider the
flux creep under the influence of a
transport current parallel to the defect
planes which creates a driving force
density
$\textbf{f}=\textbf{J}\wedge\textbf{ B}/c$
perpendicular to them. $\textbf{J}$ is the
current density. The motion of the FL
bundles under the influence of
$\textbf{f}$ occurs then by nucleation of
critical droplets in which FLs are moved
by a distance $\ell$. This droplet is a
saddle point, as usual in nucleation
phenomena. In the presence of planar
defects the energy of the nucleus has the
from
\vspace{-0.15cm}
\begin{align}\label{eq:E-nucl}
E_{nucl}\approx L^{d-2}L_yL_z\left(\frac{c_{11}\ell^2}{L^2}+\frac{\Sigma_z}{L_z}+ \frac{\Sigma_y}{L_y}-f\ell \right)
\end{align}
Here we have taken into account that the
elastic energy and the energy from the
disorder scale in the same way. The
saddle point $L_y/\Sigma_y=L_z/\Sigma_x\sim
L^2/c_{11}\sim f^{-1}$ gives for the
nonlinear resistivity in $d=3$ ($J\ll
J_D$)
\vspace{-0.15cm}
\begin{equation}\label{eq:rho(J)}
 \rho(J)\sim
 e^{-\left({J_D}/{J}\right)^{3/2}},\,\,\,\,J_D
 ={\cal C}\frac{(\Sigma_y\Sigma_z)^{2/3}( c_{11}/\ell)^{1/3}c}{B T^{2/3}}.
\end{equation}
Thus the non-linear resistivity is reduced
considerably with respect to the case of
point impurities. A similar consideration for the Bose glass  gives
$\mu=1$ which is, as far as we are aware, also a new result \cite{Blatter_93}.

To summarize
 the results obtained so far we remark that
the new phase described here is
characterized by
(i)    diverging elastic
constants $c_{44}$ and $c_{66}$ but a finite
compressibility $c_{11}$, (ii) a transverse
Meissner effect as well as  a resistance against shear deformation,  (iii) large sample to
sample fluctuations of the susceptibility,
(iv) an exponential decay of  the
structural correlations (in $d=3$) and (v)
a creep exponent $\mu=3/2$. Since the
totality of these properties  is different
from the Bragg glass or  the Bose glass,
 we will call this  new phase a \emph{
planar glass}. This phase is also
different from that found for equally
spaced defects which is incompressible
\cite{BaNe94}.

\emph{Strong disorder}.
If the disorder is strong, i.e.~if
$L_D\lesssim \ell_D$, (we ignore for the
moment the point disorder) each defect
will be completely overlapped by a FLL
plane to gain its full energy.
Integrating out the displacement
field between two adjacent defect planes
we get in $d=3$
\vspace{-0.15cm}
\begin{align}\label{}\nonumber
 \frac{\cal H}{L_zL_y}=\sum_{i=1}^{N}\Big\{\frac{c_{11}}{2}
 \frac{(u_{i+1}-u_i)^2}{\Delta x_{i+1}}-\rho_0v_{D}\sum_{n}e^{i G_D n
(x_i-u_i)}\Big\}
\end{align}
where $G_D=2\pi/\ell$, $\Delta x_{i+1}=x_{i+1}-x_{i}$ and
the sum over $i$ is over the defect
planes. 
For $v_D\to \infty$ we have
$x_i-u_i=\ell n_i$ with $n_i$ integer to
minimizes the pinning potential.
Minimizing subsequently the elastic energy
allows the exact determination of the
ground state \cite{GlNa04}:
$
 u_i^0=x_i-\ell\sum_{j=1}^i\left[{\Delta x_j}/{\ell}
 \right]_G
$, where $[x]_G$ denotes the closest integer to x. For
$\ell_D\gg \ell$,  $S_{\bf G}({\bf
r})$ is again decaying exponentially
in the $x-$direction on scale  $\ell_D$.
Considering  flux creep due to a
driving force $\textbf{f}$ perpendicular
to the defect planes
in $d=3$ we obtain
then the same form of the non-linear
resistivity Eq.~(\ref{eq:rho(J)}) as in the
case of weak disorder. This
formula applies for small currents where
droplets cover many planar defects. Thus
both weak  and strong disorder give the
same results for the correlations and the
flux creep.

\emph{Displacement parallel to the
defects, dislocations}. Next we include displacements $u_y$
parallel to the defects. In the case of
strong disorder  each defect is
occupied by a single  FL layer and hence
$u_x(x_i,y,z;n_i)=x_i-\ell n_i$, $\forall
y,z$, to maximize the pinning energy gain.
Even without point disorder we obtain then
a non-zero displacement $u_y$. This can be seen most easily
in the isotropic case where $\sigma\partial_x u_x =-
\partial_y u_y$, here $\sigma=(c_{11}-c_{66})/(c_{11}+c_{66})$ is the
Poisson number, $0<\sigma<1$
\cite{Landau_7}.
The strain  $\partial_x
u_x$ in the segment between the defects at $x_{i+1}$ and $x_i$ is
$\partial_x u_x\approx
1-\ell\Delta n_{i+1}/\Delta x_{i+1}$ where   $\Delta n_{i+1}=(n_{i+1}-n_i)$.
The difference  of the strain $\partial_yu_y$
in neighboring segments is then  $\Delta
\partial_yu_y\approx \sigma\ell[\Delta n_{i+1}/\Delta x_{i+1}-\Delta n_i/\Delta x_i]
$ which is of the order $\pm \sigma
\ell/\ell_D$. On the scale $L_y$ this
implies $\Delta u_y\sim \pm \sigma\ell
L_y/ \ell_D$. To avoid a diverging shear
energy one has to allow for dislocations
with Burgers vector \emph{parallel} to the
$y$-direction sitting at the defects.
Their distance in the $y$-direction is of
the order $\ell_D/\sigma$. Comparing the  energy of
an edge dislocation piercing the crystal
to the energy gain from the disorder 
we find that dislocations will be present if $\sigma c_{66}\ell^3\xi_c\ll\ell_D v_D$.
In general, the network of
additional FLL sheets spanned by the
dislocations will be complicated. 
The resulting state is
ordered in the sense that  $\Sigma_{y},\Sigma_z$
are non-zero 
and  hence the transverse Meissner effect
is still present.

Adding weak point impurities will further
randomly shift the positions of the
dislocation leading most likely to decay
of translational correlations in the
$yz-$plane. Since the Burgers vector of
the dislocations is always parallel to the
defects, creep in the $x$-direction is not
facilitated.   $\Sigma_y$ and $\Sigma_z$
are still both non-zero and hence we recover
the creep law Eq.~(\ref{eq:rho(J)}). To describe creep
parallel to the defects one has to take
into account the interaction between the
dislocation, a situation not considered so
far \cite{KiNoVi00}. We leave this case for further studies.
In the case of weak pinning qualitatively
the same behavior can be expected on
scales $L_x\gg L_D$, in particular if the
flow is again to the strong coupling fixed
point. If the samples exhibits orthogonal
families of (non-intersecting) defects,
long range order in the $xy-$plane is
destroyed even without point disorder on
scales larger than $L_D$. The creep is now
limited by the slowest mechanism and hence
Eq.~(\ref{eq:rho(J)}) is likely to be valid
for all current directions in the
$xy$-plane.

The authors acknowledge support from the
SFB 608 and helpful comments from  F. de la Cruz, J.
Kierfeld, V.M. Vinokur, M. Zaiser and E. Zeldov.



\begin{thebibliography}{99}
\bibitem{dispersion} The dispersion of the tilt and the compression modulus  on small length scales \cite{Blatter_93}
 is negligible for the present problem  since we are interested in the asymptotic poperties (large scales, small currents).
\bibitem{footnote0} $R_P({\mathbf u})=
n_{imp}(v_P\rho_0)^2\sum_{\mathbf{G}\neq\mathbf{0}}
e^{i\mathbf{G} \mathbf{u}}$ , $v_p$ is the strength of an  impurity and $|\mathbf G|\le 1/\xi_c$.
\end{thebibliography}
\end{document}